# Percolative nature of the transition from 60 K to 90 K–phase in $YBa_2Cu_3O_{6+\delta}$

S.Yu. Gavrilkin, O.M. Ivanenko[*], V.P. Martovitskii, K.V. Mitsen, A.Yu. Tsvetkov

*Lebedev Physical Institute, Leninskii pr., 53, 119991, Moscow, Russia*

**Abstract**

We have measured the heat capacity of $YBa_2Cu_3O_{6+\delta}$ for $0.7<\delta<0.8$ between 1.8 and 300K. It was found that doping dependences of specific heat jump $\Delta C$ and temperature of heat capacity jump $T_{\Delta C}$ contradict to the assumption of spatially homogeneous electronic density. The results suggest that the transition from 60K to 90K phase has a percolative nature and the structure of underdoped 60K –phase can be considered as array of superconducting nanoclusters embedded in the insulating matrix.

*Keywords:* Heat capacity; YBCO; Negative-U center

It is known that two ranges of superconducting state can be distinguished on the phase diagram of $YBa_2Cu_3O_{6+\delta}$ in dependence on doping $\delta$. The first one $0.8<\delta<1.0$ (with $T_c\approx 90K$) usually considered as optimal doping range and the second one, so called underdoped 60K-phase ($0.5<\delta<0.7$ with $T_c\approx 60K$). The reasons of an essential $T_c$ difference in these phases remain to be uncertain till now. Here we studied the doping dependence of electronic specific heat jump, $\Delta C_{el}$ at superconducting transition in $YBa_2Cu_3O_{6+\delta}$ at $\delta$, corresponding to the range between 60K and 90K phases.

We used ceramic $YBa_2Cu_3O_{6+\delta}$ samples synthesized by conventional solid-state reaction. The oxygen content varied by annealing of fully oxygenated samples at different temperatures in air atmosphere with further quenching into liquid nitrogen. Oxygen concentration was estimated by X-ray diffraction measurement of lattice parameters **a** and **c** with using lattice parameters dependence on $\delta$ for $YBa_2Cu_3O_{6+\delta}$ from [1]. Oxygen homogeneity over the sample estimated from reflex width was $\Delta\delta=\pm 0,003$. The heat capacity $C(T)$ measurements were carried out using a Quantum Design Physical Properties Measurement System (PPMS). The experimental results are shown in Fig.1.

Unusual behavior of doping dependence of heat capacity jump at the temperature of superconducting transition is obvious from this figure. The jump decreases twice in the narrow range $0.80<\delta<0.86$ at the same position of jump and its temperature width. At further reduction of oxygen concentration the jump shifts towards lower temperatures with simultaneous decreasing of magnitude (at nearly fixed width) and vanishes at $\delta<0.7$ (under the given experimental resolution). Such behavior is in agreement with another experimental results [2] but doubtless contradicts to the assumption of spatially homogeneous superconducting transition of the sample. In BCS model the reduction of jump magnitude at $0.80<\delta<0.86$ should mean $\gamma$ reduction. But this contradicts to the constant critical temperature in this range of $\delta$. The observed $\Delta C(T)$ features can not be explained by inhomogeneity of dopant distribution, because for this case the jump would be expected to vanish with doping reduction at the same temperature (at phase stratification) or to spread with decreasing of value that does not observed in the experiment..

[*] Corresponding author. fax: +7-495-938-2251; e-mail: ivanenko@sci.lebedev.ru.

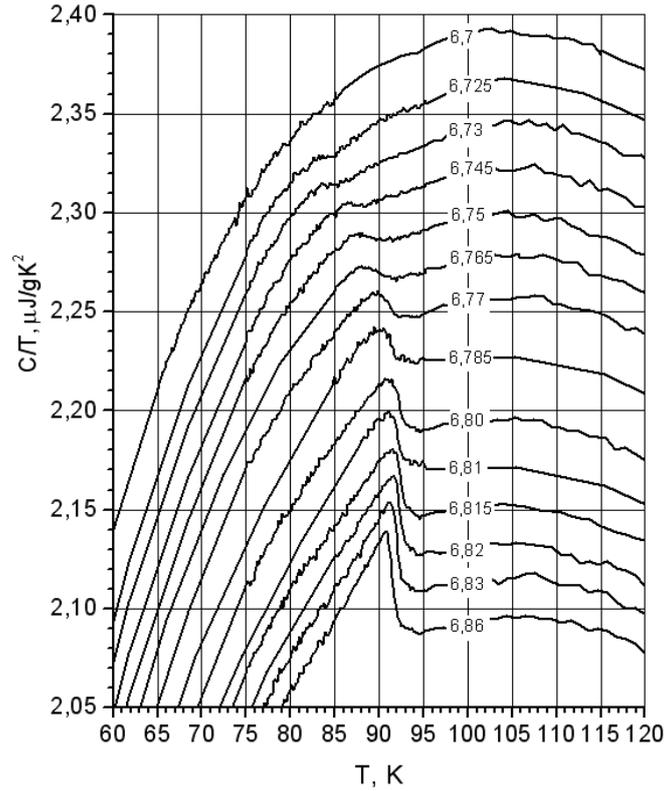

Fig. 1. The temperature dependences of heat capacity for $YBa_2Cu_3O_{6+\delta}$ with various $\delta$ (the $\delta$ values are shown in the figure). Curves for the different $\delta$ are shifted serially up.

The results was found to be in agreement with model [3] according to which at $0.8<\delta<1.0$ superconducting phase forms percolating cluster with $T_c\approx 92K$ embedded in the insulating matrix. With $\delta$ reduction the cluster power decreases and heat capacity jump decreases accordingly. When $\delta$ reduces below percolation threshold $\delta_c$ (that should correspond to reduction of a relative fraction of superconducting phase below 0.5 for two-dimensional case) a percolative cluster disintegrates on finite clusters of various sizes.

In the model [3] the interaction of valence-band electrons with so-called negative U-centres (NUC) is assumed to result in superconducting pairing. NUC is formed on pair of neighboring Cu ions in $CuO_2$ plane when three in a row oxygen sites in CuO-chain over (under) this Cu pair are occupied. Superconductivity in this model arises in 2D clusters which integrate different number of NUCs. These clusters have identical concentration of carriers, but differ in sizes. The number of Cu ions inside the NUC cluster in the $CuO_2$ plane is taken to be the size of NUC cluster $S$.

Fluctuations of NUC occupation for the finite clusters will lead to $T_c$ lowering with respect to $T_{c\infty}$ of infinite percolation cluster, for which influence of fluctuations is negligible and consequently $T_{c\infty}$ = const (see fig.5 from [3]). Assuming random distribution of oxygen in chain planes, we have calculated the statistics of NUC clusters in $CuO_2$ planes for various $\delta$ for a 500x500 Cu site lattice. The value of percolation threshold along NUC was found to be $\delta_c$=0.78. The dependences of number of clusters $N$ on their size for varied $\delta$ are shown in Fig. 2. As evident from Fig. 2 the majority of finite clusters involves 10-100 sites that corresponds to $T_c$ = (50-70) K (according to fig.5 from [3]). This is the so-called 60K-phase. As far as $T_c$ for a finite cluster goes down with $S$ reduction, the superconducting transition for the sample with given δ will happen step-by-step for different clusters at temperature lowering, starting from $T=T_{c1}(\delta)$ - the transition temperature for the cluster with maximum size $S_1(\delta)$. At $T<T_{c1}$ the volume fraction of superconducting phase increases with temperature lowering due to the superconducting transition of smaller clusters. Due to variety of cluster sizes, a jump of heat capacity below the percolation threshold rapidly disappears, spreading over a broad temperature range. Only in a narrow range just below the percolation threshold, where the existence of sufficiently large finite cluster (much exceeding all other clusters) is possible, the heat capacity displays the jump connected with superconducting transition of this cluster. In this case $\Delta C$ will be proportional to $S_1(\delta)$. Below the percolation threshold the size of the biggest finite cluster $S_1$ rapidly falls with doping reduction (see Fig. 5 from [3]), accordingly, $T_{c1}$ starts to depend on doping. It is this behavior that was observed in the experiment.

Unusual behavior of doping dependence of heat capacity jump at the temperature of superconducting transition is obvious from this figure. The jump decreases twice in the narrow range $0.80<\delta<0.86$ at the same position of jump and its temperature width. At further reduction of oxygen concentration the jump shifts towards lower temperatures with simultaneous decreasing of magnitude (at nearly fixed width) and vanishes at $\delta<0.7$ (under the given experimental resolution). Such behavior is in agreement with another experimental results [2] but doubtless contradicts to the assumption of spatially homogeneous superconducting transition of the sample. In BCS model the reduction of jump magnitude at $0.80<\delta<0.86$ should mean $\gamma$ reduction. But this contradicts to the constant critical temperature in this range of $\delta$. The observed $\Delta C(T)$ features can not be explained by inhomogeneity of dopant distribution, because for this case the jump would be expected to vanish with doping reduction at the same temperature (at phase stratification) or to spread with decreasing of value that does not observed in the experiment..

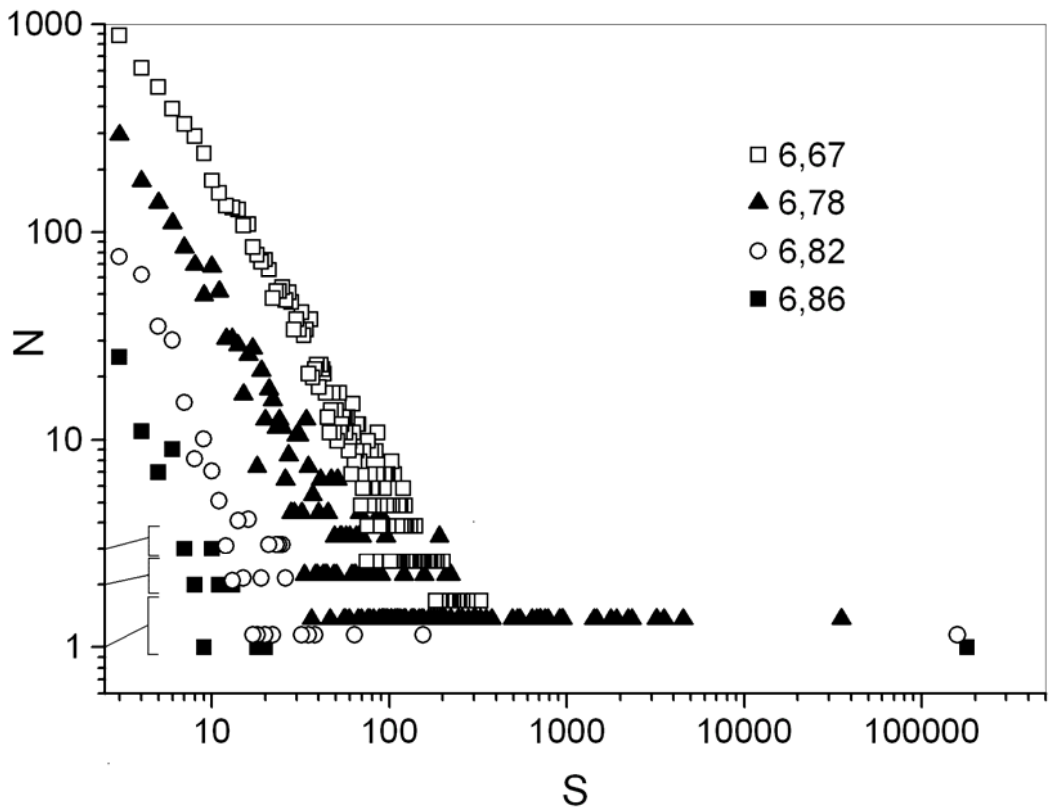

Fig. 2. The dependence of cluster number of NUC- $N$ on their size –$S$ for various $\delta$ at random occupation of oxygen sites in chains. $N(S)$ for various $\delta$ are shifted serially up. (Square brackets in the figure unify groups of clusters with $N$=1, 2, and 3)

The experimental results $\Delta C(\delta)/T$ are in a good agreement with $S_1(\delta)$ dependence obtained in [3]. The similar agreement takes place for doping dependences of temperature of heat capacity jump $T_{\Delta C}$ and temperature of superconducting transition for the biggest cluster $T_{c1}$.

In conclusion, the results give the reason to consider the underdoped 60K-phase as array of superconducting nanoclusters embedded in the insulating matrix. With doping increase these nanoclusters unite into the aggregate percolation cluster with $T_c$=92K. The fact, that the used model allows to give a qualitative and quantitative explanation of our experimental results, can be considered as argument in favor of fundamentals of the model [3].

This research was supported by the Russian Foundation for Basic Research (grant 08-02-00881)

**References**

[1] Y. Nakazawa, M. Ishikawa, Physica C, 158 (1989) 381.
[2] J.W. Loram *et al*. Phys.Rev.Lett. 71 (1993) 1740
[3] K.V. Mitsen, O.M. Ivanenko, JETP 107 (2008) 984